# Tuning the Stabilization Mechanism of Nanoparticle-Regulated Complex Fluids

*Marzieh Moradi[a], Qingwen He[a], Gerold A. Willing[a]\**

[a] Department of Chemical Engineering, University of Louisville, Louisville, KY 40292, USA

*Corresponding author. Tel.:+1-502-852-7860; fax: +1-502-852-6355.

E-mail address: gerold.willing@louisville.edu

**Abstract**

In nanoparticle haloing, charged nanoparticles have been found to enhance the stability of colloidal suspensions by forming a non-adsorbing layer surrounding neutral colloids which induces an electrostatic repulsion between them. However, there has been some debate that nanoparticles may directly deposit onto the colloidal surfaces and that the stabilization mechanism relies on nanoparticle adsorption. In this study, we have found that these two mechanisms control the stability of colloidal suspensions across a continuum over increasing nanoparticle concentrations. AFM force measurements showed that highly charged zirconia nanoparticles built up an electrostatic repulsion between negligibly charged silica surfaces, preventing them from aggregating. The follow-up adsorption measurements and force modeling indicated that minor adsorption of nanoparticles is expected at volume fractions of $10^{-5}$ to $10^{-3}$, but the amount of nanoparticle adsorption dramatically increases with increasing the nanoparticle volume fraction beyond $10^{-3}$. Based on these results, we propose that the fundamental mechanism of nanoparticle-regulated stabilization is nanoparticle haloing at low nanoparticle concentrations which transitions to adsorption at higher concentrations. Accordingly, at a nanoparticle volume fraction of around $10^{-3}$ where the transition happens, the stabilization can be influenced by both nanoparticle haloing and adsorption.

## 1. Introduction

Tuning the dispersion behavior of colloidal microspheres is important in several industrially relevant processes such as coatings, drug carriers, and ceramics. Traditional colloidal suspension stabilizations involve controlling the effective interactions through charged groups or deposition of polymer chains onto the colloidal surface [1], [2], [3]. However, these mechanisms have serious disadvantages, such as changing the thermal properties of the colloidal particles or contaminating the heat transfer media [4].

In 2001, a new stabilization technique regulated by nanoparticles had been proposed by Tohver et al [5], [6]. Their experimental system was composed of neutrally charged silica microparticles and highly charged zirconia nanoparticles. The colloidal suspension was found to be stabilized by using zirconia nanoparticles within a critical nanoparticle concentration range, and gelation was observed outside of this concentration window. They attributed the stabilization mechanism to nanoparticle haloing which is a non-adsorbing nanoparticle layer surrounding the colloidal particles that leads to an effective electrostatic repulsion between colloids to mitigate the inherent van der Waals attraction. Their follow-up study proved that zirconia nanoparticles were enriched near the surface of the colloidal silica and that the distance between the colloid and the loosely associated nanoparticle layer was approximately equal to the suspension's Debye length [7], [8]. Subsequently, by using Monte Carlo simulations, Liu and coworkers showed the fundamental basis for the formation of the halo is that nanoparticles are trapped in a shallow energy well close to the colloidal particle at low nanoparticle concentrations ($\leq 10^{-4}$) [9], [10]. Following the Tohver et al. experiments, nanoparticle haloing has been widely examined by a number of investigators [11], [12], [13], [14], [15], [16].

Hong and Willing used AFM to study the interaction between weakly charged silica surfaces in varying volume fractions of highly charged zirconia nanoparticles (through $10^{-6}$ to $10^{-3}$) [17]. The key feature of their observations was a peak at a separation distance of ~2 nm in the force profile measured at a volume fraction of $10^{-5}$, which could be attributed to a formation of nanoparticle halo with a 2 nm-separation distance from silica surface. Their result was in accordance with Tohver's observation that the haloing distance is roughly equal to the Debye length.

Besides the original silica-zirconia system, the colloidal stabilization regulated by charged nanoparticles has also been successfully applied to other colloid-nanoparticle binary suspensions [18], [19], [20]. Since this stabilization method does not rely on adsorption, it is specifically suitable in applications where using adsorbed species may hinder reactivity or availability of the surface such as in colloidal surface functionalization and ceramics processing. Examples of where nanoparticles have been used to increase the stability of colloidal particles in different industrial processes include ceramic processing [21], [22], chemical-mechanical polishing [23], development of paints [24], and drilling fluids [25].

Ji et al. [26] and Mckee et al. [27] also investigated the use of charged nanoparticles to manipulate the interaction between neutral colloids. They proposed that the stabilization of the binary suspension is caused by sufficient deposition of nanoparticles onto the colloids, leading to an increase in the effective charge density on the colloidal surfaces, and thereby enhancing the electrostatic repulsion between them. This increased repulsion did not vanish upon flushing the nanoparticles out of the system, indicating strong nanoparticle adsorption. This result challenges the potential use of highly charged nanoparticle as a tool to reversibly tailoring colloidal stability. However, their experiment focused on relatively high nanoparticle volume fractions ($\geq 10^{-3}$).

As mentioned above, some researchers have shown that in some of the systems the stabilization mechanism is based on the nanoparticle haloing while other researchers observed strong deposition of nanoparticles in their systems [5], [26]. Considering the widespread use of nanoparticles in the stabilization of colloidal particles and the contrary observations in the stabilization mechanism resulting from the different studies, it is very important to understand and determine the exact stabilization mechanism of the colloidal binary systems.

In this work, we conducted a comprehensive experimental study in a nanoparticle-regulated binary system by directly measuring the interaction force between a neutral silica microsphere and plate in nano-zirconia aqueous suspensions at varying nanoparticle concentrations. An adsorption measurement was also conducted to determine the degree of nanoparticle adsorption as a function of nanoparticle concentration. Our goal is to understand whether nanoparticle haloing or nanoparticle adsorption is responsible for the observed stabilization regulated by charged nanoparticles.

## 2. Materials and methods

### 2.1. Materials

Deionized water was obtained from a Mili-Q system (EMD Millipore, Billerica, MA), with a resistivity of 18 MΩ-cm. Nitric acid was purchased from Fisher Scientific, Pittsburgh, PA. Two microsphere-nanoparticle size ratios were investigated in this study: 100 (1000 nm vs. 10 nm) and 60 (600 nm vs. 10 nm). Uniform silica microspheres (Bangs Laboratories, Inc. Fisher, IN) with average diameters of 600 nm and 1 μm served as colloidal particles. The zirconia nanoparticles (Nyacol Nano Technologies Inc., Ashland, MA) were supplied in colloidal suspensions (pH≈3.5), with a manufacturer reported diameter of 10 nm with a Gaussian size distribution between 5 to 20 nm.

2.2. Methods

By adding an appropriate amount of DI water, the volume fractions of zirconia suspensions were prepared to be $10^{-2}$, $10^{-3}$, $10^{-4}$, $10^{-5}$ and $10^{-6}$. The binary suspensions for zeta potential measurements were prepared by adding an appropriate amount of silica microspheres to the zirconia suspensions. In the binary suspension, the volume fraction of silica microspheres was kept constantly at $10^{-3}$, whereas the volume fraction of zirconia varied from $10^{-6}$ to $10^{-2}$. The pH value of the mixture was then adjusted to 1.5 by adding nitric acid. An AB15 PLUS pH meter (Fisher Scientific, Pittsburgh, PA) was used to measure the pH values. Suspensions were then dispersed using ultrasonic dismembrator (Fisher Scientific, Pittsburgh, PA). Both the zeta potentials of the silica-zirconia mixtures and the particle sizes were analyzed by a 90 Plus-Zeta particle size analyzer (Brookhaven Instruments, Holtsville, NY). For measuring the zeta potential, about 1.5 ml of the binary suspension was placed in a cuvette and the electrode was inserted in it carefully to prevent any spillage. When the designated set of runs was completed, the mean values and uncertainties for the measured zeta potentials were displayed.

Force measurements were conducted between a silica microsphere and plate by an XE-100 AFM (Park Systems, Santa Clara, CA) at a scan rate of 100 nm/s in several different zirconia nanoparticle suspensions that were contained in a Petri dish. Direct measurement of the interaction force between two silica spherical particles is obviously more practical for investigating nanoparticle haloing mechanisms and directly comparing with bulk behavior, it is currently difficult to align a sphere coaxially with another sphere between smaller spherical particles (~1 μm). Because of this, the initial measurements discussed in this work were conducted between a pre-attached microsphere and a flat substrate. The result can be extended to two spheres since the separation distances are less than the radii of the curvature of microparticles and so the spherical microparticles may be assumed to be flat plates. As a matter of fact, due to the small thickness of the double layer compared to the size of the microparticles, the interaction between double layers on spherical particles may be assumed to be made up of contributions from infinitesimally small parallel flat plates between flat surfaces [28]. The silica microspheres with diameters of 600 nm and 1 μm were pre-attached on V-shaped silicon nitride cantilevers (NOVASCAN, Ames, IA, spring constant≈0.15 N/m). A silica circular plate with root mean square surface roughness of < 2 nm (height of 1/16 in., diameter of 1/2 in., Quartz Scientific, Fairport Harbor, OH) served as the flat substrate. Before the experiments, the silica plate was initially sonic cleaned for ten minutes, and then washed three times with anhydrous ethanol followed by DI water.

The amount of nanoparticle adsorption onto the silica plate was measured by scanning the surface with an FEI Nova 600 scanning electron microscope (SEM). For this purpose, dried silica plates were first fully placed in a pH-adjusted solution and then in a nanoparticle suspension for 30 min. The silica plate was then removed from the nanoparticle suspension and soaked in a pH-adjusted solution to remove the non-adsorbed nanoparticles from the plate surfaces. Then, the plate was dried under a laminar hood and taken to the SEM for imaging.

## 3. Results and discussion

3.1. Zeta potentials

At pH 1.5, the silica suspension has a negligible surface potential of +1mV [5] so that the electrostatic repulsion between the colloids is minimized. Under such conditions, the van der Waals attraction is the only operating force, allowing all silica colloids to aggregate and precipitate out of suspension. Zirconia nanoparticles are highly charged at this pH value, having a consistent zeta potential of 70 mV with zirconia volume fraction varying from $10^{-6}$ to $10^{-2}$ [5]. The effective zeta potentials of the binary system of silica colloids associated with charged nanoparticles at pH 1.5 are presented in Table 1, showing that the effective zeta potential of the binary mixture increases with increasing the nanoparticle concentration. When the nanoparticle concentration is as low as $10^{-6}$, the effective zeta potential is around 20 mV, indicating an unstable state of colloidal suspension. As the nanoparticle volume fraction grows to $10^{-2}$, the zeta potential increases to approximately 60 mV, suggesting a good stability.

**Table 1.** Effective zeta potentials of the microsphere-nanoparticle binary suspensions at varying nanoparticle volume fractions ($\Phi_{silica}= 10^{-3}$; $\Phi_{zirconia}= 10^{-6}$ -$10^{-2}$).

| Size Ratio | Zeta potential (mV) | | | | |
|---|---|---|---|---|---|
| | $10^{-6}$ | $10^{-5}$ | $10^{-4}$ | $10^{-3}$ | $10^{-2}$ |
| 100 | 18.4±0.6 | 30.0±1.2 | 38.0±1.7 | 45.0±1.7 | 60.4±2.9 |
| 60 | 21.7±0.9 | 30.2±1.2 | 40.4±1.7 | 55.2±2.0 | 65.1±2.7 |

3.2. Force profiles in presence of nanoparticles

The measured interaction forces between a weakly charged silica microsphere and plate in a highly charged zirconia nanoparticle suspension are shown in Fig. 1. Each force curve is the average of 20 repeated force curves obtained from different locations under the same conditions. Slightly different force curves obtained each time, which may be caused by the change of the nanoparticle number density in the sphere-flat gap due to the dynamic nanoparticle diffusion driven by the probe motion. These changes were less obvious at nanoparticle volume fractions higher than $10^{-5}$ since there are far more nanoparticles in the gap at higher concentrations. Due to the variable nanoparticle size distribution, slightly larger nanoparticles would begin to force out of the sphere-flat gap when the probe approaches the surface. In the opposite case, smaller nanoparticles are drawn into the gap when the probe retracts from the surface. This may cause a variation in the number of the nanoparticles in the gap which leads to variances in the zeta potential and ultimately to a change in the resulting force profile. Detailed statistical analysis of the AFM force measurements can be found in our previous work [17]. As seen in Fig. 1a, at a size ratio of 100, the interaction between a silica microsphere and plate in nanoparticle suspensions at pH 1.5 is sensitively dependent upon the nanoparticle volume fraction. The interaction between silica surfaces is purely attractive when no nanoparticles are added, or the nanoparticle volume fraction is as low as $10^{-6}$. But it becomes completely repulsive as nanoparticle volume fraction is increased to $10^{-5}$, and the repulsion becomes stronger as the nanoparticle volume fraction increases. When the interaction between colloidal surfaces is dominated by repulsion with volume fraction $\geq 10^{-5}$, colloidal particles are prevented from aggregation, thus leading to a state of colloidal system stability. This result demonstrates that highly charged nanoparticles produce an effective repulsive force between the neutral colloidal particles. A similar result is observed at the size ratio of 60 (Fig. 1b).

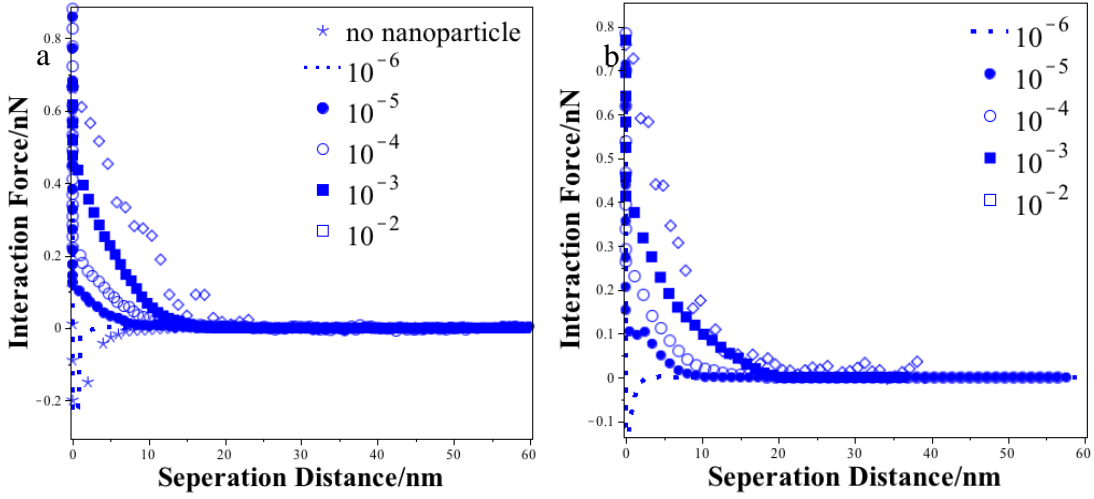

**Fig. 1.** Force profiles between a silica microsphere and plate in varying volume fractions of zirconia nanoparticles at size ratio of (a) 100; (b) 60.

### 3.3. Decay Length Fitting Model

In our previous work, a Decay Length Fitting Model (DLFM) was developed to simply estimate the interaction between colloidal surfaces in the presence of nanoparticles [29]. In this model the total interaction is expressed as the combination of the van der Waals force, electric repulsive force and depletion force: $F_{total}(D) = F_{vdW} + F_{electro} + F_{depletion}$ (1). The approximated equation for each of the forces is as follows:

$$F_{vdW}(D) = -\frac{AR}{6D^2} \quad (2)$$

$$F_{electro}(D) = +\kappa R \varepsilon_0 \varepsilon_r \pi \Psi_{effective}^2 \exp(-\kappa D) \quad (3)$$

$$F_{depletion}(D) = \begin{cases} -2\rho_\infty kT\pi \left[a^2 + 2aR - RD - \frac{D^2}{4}\right] & for\ 0 \leq D \leq 2a \\ 0 & for\quad D > 2a \end{cases} \quad (4)$$

Where $D$ is the separation distance of the closest approach between the sphere and the plate, $R$ is the radius of the microparticle and $A$ is the Hamaker constant of silica ($0.8\times10^{-20}$ J) [30]. $\varepsilon_0$ is the vacuum permittivity, $\varepsilon_r$ is the dielectric constant, $\kappa$ is the reciprocal of the Debye length, $\Psi_{effective}$ is zeta potential of the binary mixture which is presented in Table 1, $a$ is the nanoparticle size, $\rho_\infty$ is the bulk number density, and $kT$ is the thermal energy. The van der Waals attraction between a silica sphere and plate is calculated using the simplified expression of Hamaker when the colloidal sizes are sufficiently large compared to the distance between them [27]. The electrostatic repulsion is calculated using the Hogg-Healy-Fuerstenau (HHF) formula [28], which is well known to calculate the double layer interactions at constant surface potential between dissimilar surfaces and has been utilized to study mechanisms of nanoparticle haloing in several previously reported works [9], [10]. We assume that the effective zeta potential is the same for both silica sphere and plate, and a continuum assumption is made for the overlapping of the effective double layers as the silica sphere approaches the plate. The depletion force is estimated using Piech and Walz's approximation [31]. Despite the fact that the depletion force was neglected in our original work with the DLFM [29], the depletion force is taken into account in this study due to the higher nanoparticle concentrations dealt with herein.

According to previous nanoparticle-haloing studies [9], [10], highly charged nanoparticles are trapped in a shallow energy well close to the neutral colloidal particle, so that nanoparticles would be easily squeezed out of the gap region between colloidal surfaces while they are approaching. Therefore, in the DLFM we assume the van der Waals attraction between silica surfaces is barely affected by zirconia nanoparticles, and is calculated by using the regular Hamaker's model without taking the impact of nanoparticle adsorption into account.

It should be noted that the term $\kappa$ in Eq. (3) stands for the decay length of the colloid-nanoparticle mixture. Under the conditions of the nanoparticle halo, highly charged nanoparticles would segregate to each negligibly charged colloidal surface, forming a loose nanoparticle layer located a small distance away from the colloidal surface [5], [6], [7], [29]. This means the gap between the microsphere surface and its effective charge plane is affected by the separation between the microsphere and the nanoparticle halo as well as the effective surface charge density of colloidal particles. As a result, the thickness of the charge layer cannot be calculated using the regular Debye length equation. Therefore, $\kappa$ is taken as a variable and is to be determined by fitting Eq. (1) to the experimental interactions. The total interaction equation with decay length as the fitting parameter is named DLFM.

We substituted the measured effective zeta potential values (as shown in Table 1) into Eq. (1) and then adjusted the value of the decay length until the theoretical and experimental force curves matched. Fitting values of the decay length as a function of nanoparticle volume fraction are presented in Table 2. The final fitting results at different size ratios are shown in Fig. 2, with the solid lines representing the total interaction force as calculated by the DLFM.

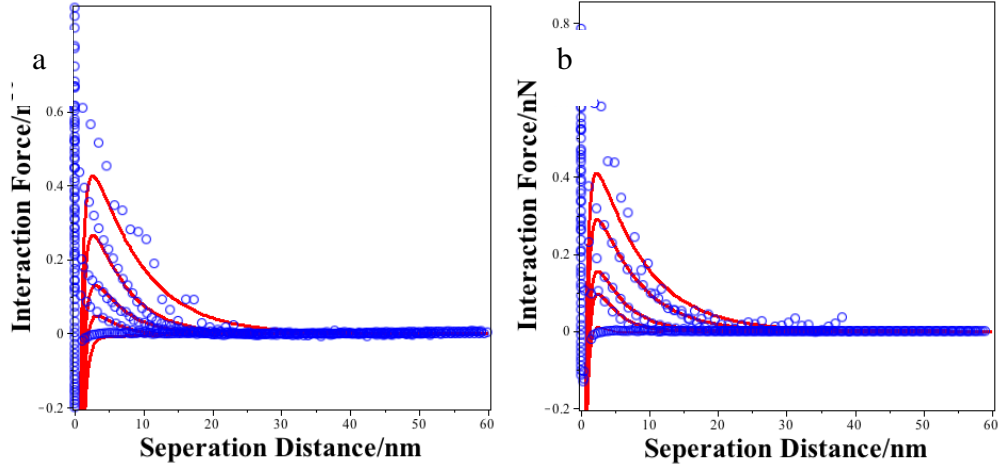

**Fig. 2.** Experimental force curves and fitting curves at size ratio of (a) 100; (b) 60. The solid lines represent the total interaction force calculated by the DLFM.

As seen in Fig. 2, the interaction calculated by the DLFM shows a good agreement with the experimental data as the nanoparticle volume fraction increases from $10^{-6}$ to $10^{-3}$, indicating the interaction between weakly charged colloidal surfaces in the presence of highly charged nanoparticles is mainly composed of an effective electrostatic repulsion, van der Waals attraction and depletion force at these concentrations. Despite the fact that the silica particles are weakly charged at pH 1.5, an effective repulsion arises from the addition of highly charged nanoparticles, demonstrating that these nanoparticles can induce an effective plane of charge around the neutral colloids. Moreover, the good fit of the DLFM from $10^{-6}$ to $10^{-3}$ implies that only a minor nanoparticle adsorption is expected at these nanoparticle concentrations, so that the van der Waals attraction between the colloids will not be shielded by an adsorbing nanoparticle layer. It should be noted that as predicted by the DLFM, the van der Waals forces take over as the separation distances become very small. However, AFM force measurements are dynamic measurements which properly approximate equilibrium measurements at higher separation distances and low approach speeds. But at very small separation distances, the spring constant is not strong enough to accurately measure the strong van der Waals forces which results in a jump to contact between the sphere and the flat.

**Table 2.** Fitting results of decay lengths (nm) at varying nanoparticle volume fractions using DLFM model.

|  | Volume Fraction | | | |
| --- | --- | --- | --- | --- |
| Size Ratio | $10^{-6}$ | $10^{-5}$ | $10^{-4}$ | $10^{-3}$ |
| 100 | 2.0±0.1 | 3.3±0.2 | 4.2±0.3 | 6.2±0.4 |
| 60 | 2.5±0.1 | 3.0±0.2 | 4.0±0.2 | 6.5±0.4 |

As shown in Table 2, the fitting results of the decay length show a successive increase with increasing nanoparticle concentration. Decay length uncertainties were propagated from zeta potential and nanoparticle size uncertainties. It should be noted that at pH 1.5, the Debye length in a pure nitric acid solution is only 1.7 nm [27] which is smaller than the decay lengths that are found in any of the silica-zirconia binary systems.

Since the charge layer of the colloidal particle is induced by the nanoparticles, two factors may impact the decay length of the binary mixture. The first is the non-zero separation between the colloidal surface and nanoparticle halo, which slightly shifts the effective charge layer away from the silica surface. The second is the nanoparticle density near the colloidal particle. When the number of nanoparticles associated with a halo increases, the effective charge density within the nanoparticle layer will increase which appears in the experimental force curves and the model as an expansion of charge layer around the colloidal surface. Due to the combination of these two factors, the decay screening length of the binary mixture is larger than the Debye length in a pure acid solution and it slightly increases with the increasing nanoparticle concentration.

It should be noted that the DLFM tended to underestimate the interaction at separation distances <10 nm at a nanoparticle volume fraction of $10^{-2}$. Since the effect of nanoparticle adsorption on the interaction force between colloidal surfaces is ignored in this model, the failure of DLFM at high nanoparticle concentration is possibly due to an alteration of the van der Waals attraction between colloidal surfaces caused by a high level of nanoparticle adsorption.

3.4. Modified Decay Length Fitting Model

If the level of nanoparticle adsorption is sufficiently high, the adsorbed layer of nanoparticles would alter the physicochemical properties of the silica surfaces (as schematically shown in Fig. 3b), thus perturbing the van der Waals attraction between them. As a result, the van der Waals equation used in DLFM may no longer remain valid at high nanoparticle volume fractions. This may explain the deviation of DLFM observed at a volume fraction of $10^{-2}$, and thus the van der Waals formula must be modified.

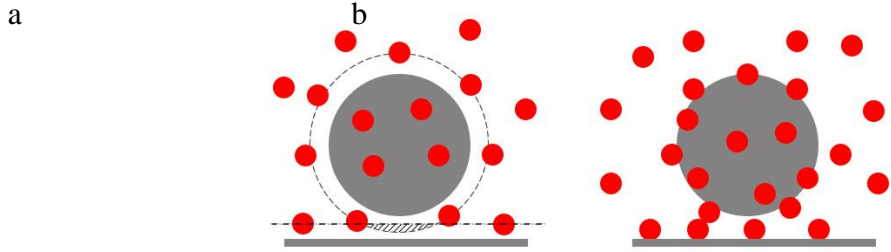

**Fig. 3.** Schematic diagram of the interactions between a probe sphere and a flat surface induced by a) nanoparticle halo; b) adsorption.

Vold had developed an approach to estimate the van der Waals attraction between spherical particles with an adsorbed layer of thickness δ [32]. Based on his formula, at $D/a_{micro} \ll 1$ and $\delta/a_{micro} \ll 1$, the interaction between a microsphere and a plate is derived as [33], [34]:

$$F_{adsorbed\_vdW}(D) = -\frac{1}{12}\left((A_w^{\frac{1}{2}} - A_{zir}^{\frac{1}{2}})^2 \frac{a_{micro}+\delta}{D} + (A_{zir}^{\frac{1}{2}} - A_{si}^{\frac{1}{2}})^2 \frac{a_{micro}}{D+2\delta} + (A_w^{\frac{1}{2}} - A_{zir}^{\frac{1}{2}})(A_{zir}^{\frac{1}{2}} - A_{si}^{\frac{1}{2}}) * \frac{4a_{micro}(a_{micro}+\delta)}{(2a_{micro}+\delta)(D+\delta)}\right) (5)$$

Where $A_w$, $A_{zir}$, and $A_{si}$ are the Hamaker constants of water, zirconia nanoparticles and silica, respectively. $A_w=3.7\times 10^{-20}$J, $A_{zir}=6\times 10^{-20}$J, $A_{si}=0.8\times 10^{-20}$J [35], [36], [37]. The thickness of the adsorbed layer is taken as 10 nm, which is the average diameter of the zirconia nanoparticles.

The total interaction equation under high-adsorption conditions is then calculated using the modified Decay Length Fitting Model (modified DLFM) as Eq. (6):

$$F_{adsorbed\_total}(D) = F_{adsorbed\_vdW} + F_{electro} + F_{depletion} (6)$$

Subsequently, the total interaction force between a silica microsphere and plate in suspensions with zirconia volume fractions of $10^{-2}$ and $10^{-4}$ is recalculated by using the revised total interaction equation (6); the fitting curves calculated using the original DLFM are also graphed for comparison.

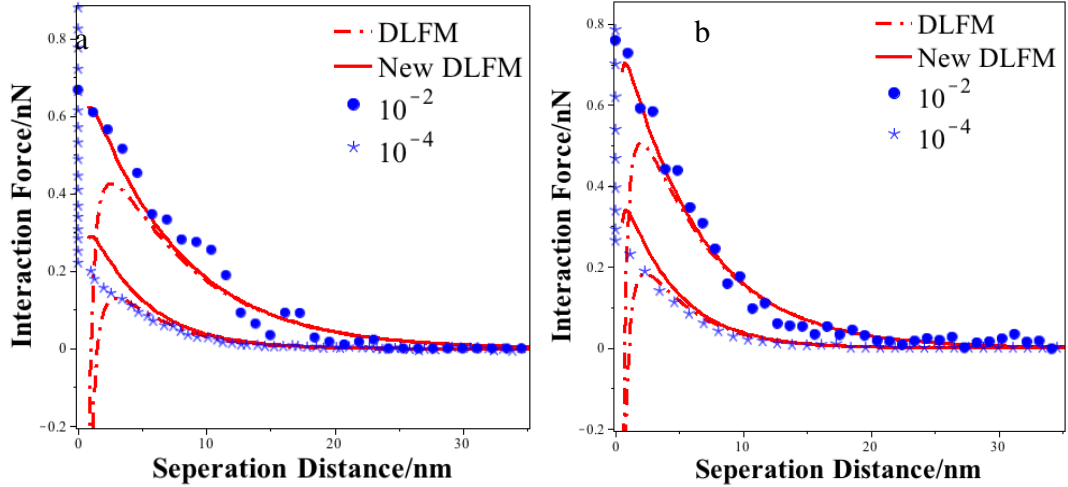

**Fig. 4.** Force curves measured at volume fraction $10^{-2}$ and $10^{-4}$, and fitting results using the modified DLFM at size ratio of a) 100; b) 60. Solid lines represent the modified DLFM fitting curve; dashed lines represent the original DLFM fitting curve.

As seen in Fig. 4, the interaction estimated by the modified DLFM has a much better fit to the experimental data at nanoparticle volume fraction of $10^{-2}$. This result confirms that nanoparticles adsorbed onto the colloidal surfaces at high nanoparticle concentration are directly impacting the interaction between the larger particles. This adsorbing layer weakens the van der Waals attraction between the silica surfaces leading to a relative strengthening of the total repulsive interaction as a result. Under such conditions, the effective charge layer of the colloidal particle is primarily produced by the adsorbing nanoparticles. However, the failure of the modified DLFM at nanoparticle volume fraction of $10^{-4}$ shows that at low nanoparticle concentrations, the effect of any adsorbed nanoparticles is minimal, and the nanoparticles are more likely haloing around the silica spheres with a non-zero separation distance between them. Therefore, final values of decay length at $10^{-2}$ was determined by the modified DLFM (Table 3) while decay length at $10^{-6}$ to $10^{-3}$ were determined using DLFM.

**Table 3.** Fitting results of decay lengths (nm) at $10^{-2}$ nanoparticle volume fraction using the modified DLFM model.

| Size Ratio | Decay Length |
|---|---|
| 100 | 6.0±0.4 |
| 60 | 6.1±0.4 |

Comparing decay length results in Table 3 to those in Table 2, decay length at the volume fraction of $10^{-2}$ remains nearly the same as that at $10^{-3}$. The possible reason is that with a high level of adsorption the physical distance between halo and the colloidal particle will gradually be filled with adsorbing nanoparticles and the effective charge layer will collapse onto the colloidal surface. Thus, even though the effective charge density will increase at $10^{-2}$ nanoparticle concentration, the overall decay length will not increase when compared to that at $10^{-3}$.

3.5. Nanoparticle adsorption measurements

To elucidate the level of nanoparticle adsorption as a function of nanoparticle concentration, the surfaces of silica plates that had been immersed in different nanoparticle volume fraction at pH 1.5 were scanned by SEM after 30 min of deposition. Several locations were scanned for each silica plate. The number of nanoparticles and average surface coverage (θ%) at different nanoparticle volume fractions was estimated by ImageJ (a Java-based image processing program) based on these SEM images. For example, Fig. 5a presents the SEM picture obtained on a silica plate that had been immersed in a zirconia nanoparticle suspension at a volume fraction of $10^{-4}$. Nanoparticles were highlighted by using ImageJ (Fig. 5b), and the calculated coverage fraction was 2.6±0.1%. The coverage fractions for $10^{-3}$ and $10^{-2}$ nanoparticle volume fractions were 18±1 and 50±3, respectively. Note that these values are based on the chosen cutoffs in ImageJ. This is, to some extent, dependent on the eye of the observer which adds some uncertainty to the calculated coverage fractions based on the maximum and minimum cutoffs that were selected.

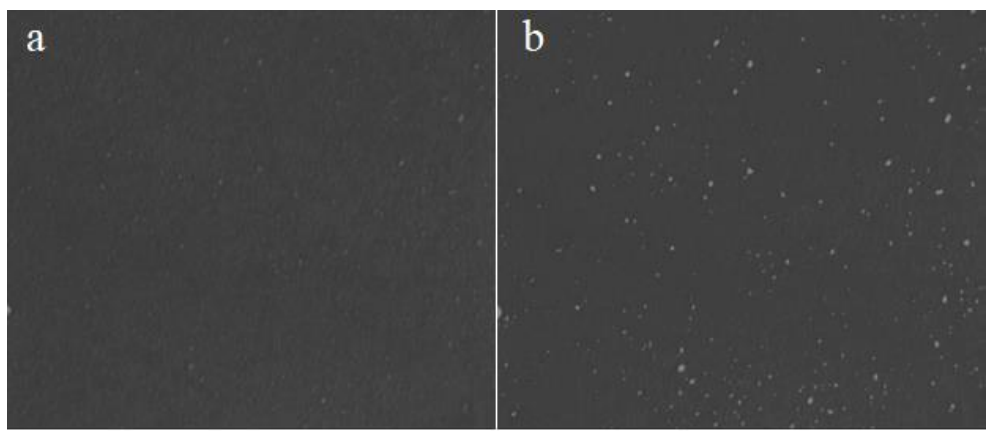

**Fig. 5.** a) SEM image obtained on silica plate that had been immersed in $10^{-4}$ nanoparticle solution (with a size ratio of 100) for 30 min at pH 1.5. b) Zirconia nanoparticles are highlighted by ImageJ

3.6. Adsorption isotherm

The experimentally determined adsorption of nanoparticles onto the silica plate was then tested graphically by fitting with various isotherms. Fig. 6. shows the experimental adsorption results together with a Langmuir adsorption isotherm. The model fittings are in good agreement with experimental data.

The Langmuir isotherm equation is mathematically expressed as follows [38, 39]:

$$q_e = \frac{K q_m C_e}{1 + K C_e} \quad (7)$$

Where $q_e$ is the amount of nanoparticle adsorbed at equilibrium (mg/g), K is the Langmuir constant, $q_m$ is the theoretical maximum adsorption capacity (mg/g), and $C_e$ is the equilibrium concentration of nanoparticle suspension (mg/L). We converted $q_e$ and $C_e$ to surface coverage ratio ($\theta_e$) and volume fraction ($C_{ve}$), respectively. Eq. (7) is rewritten as Eq. (8).

$$\theta_e = \frac{K \theta_m C_{ve}}{1 + K C_{ve}} \quad (8)$$

Where $\theta_m$ is the theoretical maximum surface coverage ratio. The linear form of Eq. (8) is rearranged as Eq. (9).

$$\frac{C_{ve}}{\theta_e} = \frac{1}{K \theta_m} + \frac{C_{ve}}{\theta_m} \quad (9)$$

The isotherm parameters obtained from the linear plot of $C_{ve}/\theta_e$ versus $C_{ve}$ (Fig. 6b) are listed in Table 4.

**Table 4.** Parameters of the best fit of the data in Fig. 6b.

| Parameter | $\theta_m$(%) | $K$ | $R^2$ |
|---|---|---|---|
| Value | 59 | 418 | 0.97 |

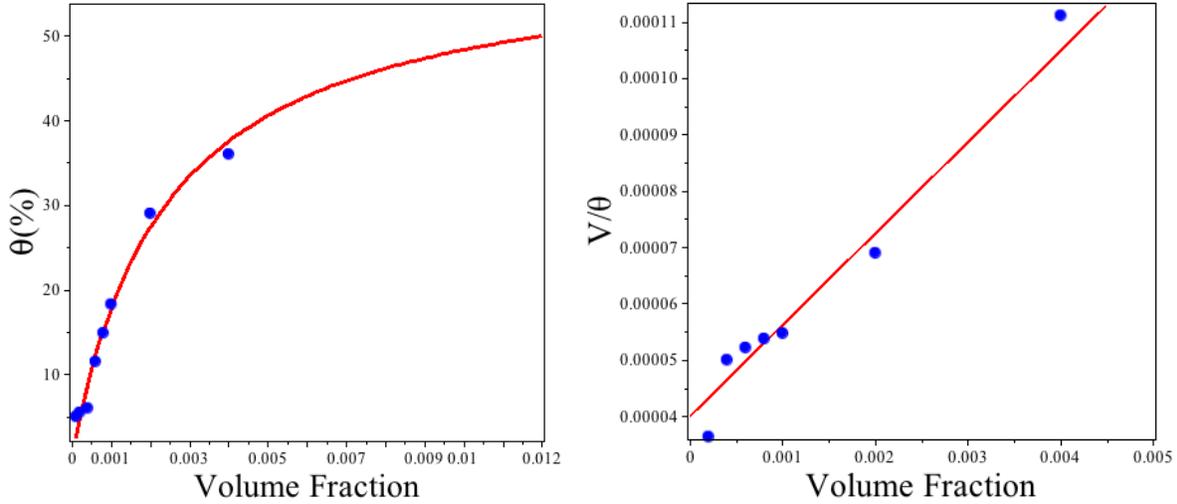

**Fig. 6.** (a) Langmuir adsorption isotherm obtained for binary suspensions with size ratio of 100 using surface coverage values (calculated from ImageJ using SEM images at each concentration) (b) Linear form of the Langmuir isotherm by plotting volume fraction/surface coverage. The solid curve corresponds to the Langmuir isotherm fit and symbols represent the experimental data calculated from ImageJ.

The Langmuir adsorption isotherm assumes that energy of adsorption is constant and adsorption process is monolayer over the homogeneous adsorbent surface [40]. As seen in Fig. 6a, the nanoparticle adsorption is very small with $\theta$ lower than 4% when nanoparticle volume fraction is less than $0.5 \times 10^{-3}$. But after that, $\theta$ sensitively increases with the increasing nanoparticle volume fraction. $\theta$ grows to approximately 18% at volume fraction of $10^{-3}$, and continually increases to about 50% at volume fraction of $10^{-2}$. It should be noted that this is still less than the theoretical maximum surface coverage of 59% that is predicted by the Langmuir Isotherm under these conditions.

3.7. Stabilization mechanism at different nanoparticle volume fractions

If we assume that nanoparticle haloing does not exist and nanoparticles are adsorbed on the surface, the surface charge of silica colloid will merely rely on nanoparticle adsorption. The surface coverage $\theta$ that is required to provide this increased surface charge at different nanoparticle volume fractions has been approximated (Results are shown in Table 5; size ratio = 100). For this purpose, the surface charge density ($q$) of a silica particle was calculated using the measured zeta potential of the binary mixture by the Loeb equation [41].

$$q = \frac{\varepsilon_0 \varepsilon_r kT}{ez} \kappa [2 \sinh\left(\frac{ez\Psi_{effective}}{2kT}\right) + \frac{4}{\kappa a} \tanh\left(\frac{ez\Psi_{effective}}{4kT}\right)] \quad (10)$$

Where $z$ is the charge of the ions in solution, $e$ is the electron charge, $\kappa$ is the Debye length of a regular silica suspension at pH 1.5 (1.7 nm), and $a$ is the particle radius.

The total charge Q for a single nanoparticle is $7.5\times10^{-18}$ C ($Q=4\pi a^2 q$, $\Psi_{zirconia}=70$ mV). We assumed that energy of adsorption is constant and adsorption layer is monolayer as required by the Langmuir Isotherm [40]. The number of adsorbed nanoparticles needed to provide the surface charge of the silica was calculated by dividing the total charge of the silica by the total charge of the nanoparticles. Then, surface coverage fraction θ required to provide the observed surface charge density was estimated. These required surface coverage values were then compared to the experimental surface coverage values measured by fitting Langmuir isotherm to the experimental data. Results are shown in Table 5.

**Table 5.** The surface coverage fraction required to provide the effective surface charge measured in the experiments (size ratio=100).

|  | Volume fraction | | |
| --- | --- | --- | --- |
|  | $10^{-4}$ | $10^{-3}$ | $10^{-2}$ |
| Zeta Potential (mV) | 38 | 45 | 60 |
| $q_{silica}$ (C/m$^2$) | $1.73\times10^{-2}$ | $2.12\times10^{-2}$ | $3.11\times10^{-2}$ |
| Required $\theta$ (%) | 23 | 28 | 42 |
| Measured $\theta$ (%) | 2.6±0.1 | 18±1 | 50±3 |

As shown in this table, at $10^{-4}$ volume fraction, a surface coverage of 23% is needed to induce the observed charge buildup, which is much larger than the experimentally determined coverage (2.6%). This result suggests that if nanoparticles do not accumulate near the colloidal particle, the surface charge induced by the adsorbing nanoparticles alone will not be able to afford the electrostatic repulsion observed at volume fraction of $10^{-4}$ due to a minor level of surface coverage. Therefore, it can be reasoned that at such low nanoparticle concentrations the effective surface charge that stabilizes the colloidal system is mainly built up by a non-adsorbing nanoparticle layer that is surrounding the neutral colloidal particles. In other words, at this low nanoparticle concentration the surface charge of the colloidal particle is mainly produced by nanoparticles that are not directly adsorbed onto the colloidal surface and colloidal stabilization is dominated by nanoparticle haloing. This result is in accordance with Tohver's finding which shows that the adsorption of nanoparticles on a colloidal particle is negligible at lower nanoparticle volume fractions and nanoparticles are found enriched at a distance of 2 nm away from the colloidal surface [8], [7].

At volume fraction of $10^{-3}$ the difference between the required and experimental $\theta$ has narrowed. This result suggests that in addition to the nanoparticle halo, the adsorbing nanoparticles will also contribute to the increased effective surface charge, and their contribution has gradually increased with the increasing surface coverage. Finally, at volume fraction of $10^{-2}$ the estimated surface coverage fraction is in good agreement with the experimental value indicating that the effective surface charge observed at high nanoparticle concentrations is primarily provided by the adsorbing nanoparticle layer. Thus, the colloidal stabilization at high nanoparticle concentration is dominated by nanoparticle adsorption.

The development of the nanoparticle layer with the increasing nanoparticle concentration is schematically described in Fig. 7. This figure shows that at low nanoparticle volume fractions ($10^{-5}$ to $10^{-4}$), nanoparticles form a non-adsorbing nanoparticle layer around the colloidal particle, increasing the effective surface charge and producing an electrostatic repulsion that mitigates the inherent van der Waals force between them. At volume fractions around $10^{-3}$, with some extent of nanoparticle adsorption, a growing portion of the effective surface charge is attributed to the adsorbing nanoparticles, leading to the stabilization mechanism being influenced by both nanoparticle haloing and adsorption. Finally, at high nanoparticle volume fractions ($\sim 10^{-2}$) the colloidal surface is significantly occupied by the adsorbing nanoparticles, and instead of nanoparticle haloing, the increased surface charge is primarily induced by nanoparticles that are directly adsorbed onto the silica surfaces.

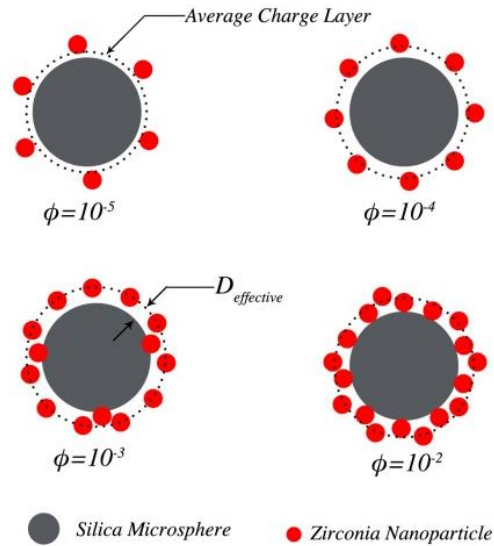

**Fig. 7.** Schematic description of the development of nanoparticle layer with the increasing nanoparticle concentration.

Despite the fact that nanoparticles are capable of stabilizing colloidal suspensions based on either the nanoparticle haloing or adsorption mechanism, it is of great importance to distinguish the working conditions of the stabilization. For example, if nanoparticle concentration is controlled within the nanoparticle haloing zone, the stabilization will not rely on adsorption, making it suitable in applications where using an adsorbed species may hinder reactivity or availability of the colloidal surface such as in colloidal surface functionalization and ceramics processing. Further study is needed to elucidate if this transition between "haloing" and "adsorption" is sensitive to the suspension properties, including nanoparticle size and surface charge. Additionally, the starting point of transition needs to be more accurately determined. Finally, a more thorough study of the transition region needs to be conducted to determine how the stabilization mechanism shifts from haloing to adsorption.

## 4. Conclusion

We have performed a comprehensive experimental investigation on the interaction between neutral colloidal surfaces in highly charged nanoparticle aqueous solutions. It is found that the silica-zirconia binary suspension system could be stabilized by highly charged nanoparticles at volume fractions ranging from $10^{-5}$ to $10^{-2}$. A subsequent adsorption isotherm study and force modeling showed that at low nanoparticle volume fraction ($10^{-5}$ to $10^{-4}$), nanoparticles form a non-adsorbing nanoparticle layer around neutral colloidal particle, which presents as an effective surface charge and produces an electrostatic repulsion that mitigates the inherent van der Waals attraction between them. In this region, there is minor amount of adsorption and colloidal stabilization is dominated by nanoparticle haloing. At high nanoparticle volume fractions (~$10^{-2}$) the colloidal surface is significantly occupied by the adsorbing nanoparticles, and instead of nanoparticle haloing, the increased surface charge is primarily induced by nanoparticles that are directly adsorbed onto the silica surfaces; Therefore, adsorption will overcome nanoparticle haloing in this region. There is a transition region around a nanoparticle volume fraction of $10^{-3}$, within which the stabilization mechanism can be influenced by both nanoparticle haloing and adsorption.

Our study suggests when using highly charged nanoparticles to stabilize colloidal suspensions, the two fundamental mechanisms of nanoparticle haloing and adsorption are not mutually exclusive. They work across the continuum to regulate the stability of colloidal suspensions over increasing nanoparticle concentrations. Depending on the ultimate application of the colloids, the primary mechanism can be controlled by simply tuning the nanoparticle concentrations. Scattering spectroscopic techniques such as in situ scattering should be done to further study the model accuracy and precision. Further studies are also needed to fully elucidate the role of nanoparticle size and charge on the transition from stabilization by nanoparticle haloing to nanoparticle adsorption and observe how the nanoparticles behave during the transition process.

**Acknowledgement**

The authors would like to acknowledge NASA EPSCoR (Grant No. NNX14AN28A) for research assistantship support for M. Moradi.

**Data Availability**

The raw data required to reproduce these findings are available and can be requested from the authors. The processed data required to reproduce these findings are available and can be requested from the authors.

**Conflicts of interest**

The authors declare that they have no conflict of interest.